\documentclass[rmp,showpacs,twocolumn,superscriptaddress]{revtex4}

 \usepackage{graphicx}

\usepackage{amssymb}

\begin{document}


\title{Magnetic properties of antiferromagnetically coupled CoFeB/Ru/CoFeB}

\author{N. Wiese}
\email[Electronic Mail: ]{mail@nilswiese.de}

\affiliation{Siemens AG, Corporate Technology, Paul-Gossen-Str. 100,
91052 Erlangen, Germany} \affiliation{University of Bielefeld, Nano
Device Group,
 Universit\"atsstr. 25, 33615 Bielefeld, Germany}

\author{T. Dimopoulos}
\affiliation{Siemens AG, Corporate Technology, Paul-Gossen-Str. 100,
91052 Erlangen, Germany}

\author{M. R\"uhrig}
\affiliation{Siemens AG, Corporate Technology, Paul-Gossen-Str. 100,
91052 Erlangen, Germany}

\author{J. Wecker}
\affiliation{Siemens AG, Corporate Technology, Paul-Gossen-Str. 100,
91052 Erlangen, Germany}

\author{G. Reiss}
\affiliation{University of Bielefeld, Nano Device Group,
Universit\"atsstr. 25, 33615 Bielefeld, Germany}

\date{\today}




\begin{abstract}
This work reports on the thermal stability of two amorphous CoFeB layers coupled antiferromagnetically via a thin Ru interlayer. The saturation field of the artificial ferrimagnet which is determined by the coupling, $J$, is almost independent on the annealing temperature up to more than $300^{\circ}$C. An annealing at more than $325^{\circ}$C significantly increases the coercivity, H$_\mathbf{c}$, indicating the onset of crystallization.
\end{abstract}


\maketitle


\section{Introduction} \label{intro}
Magnetic tunnel junctions (MTJ) have gained considerable interest in recent years due to their high potential as sensor elements \cite{Berg99} and as programmable resistance in data storage (MRAM) \cite{Gallagher97} or even data processing \cite{Richter02a}. The basic design of a spin valve consists of a hard magnetic reference electrode separated from the soft magnetic sense or storage layer by a tunnel barrier like Al$_2$O$_3$. The reference layer usually is an artificial ferrimagnet (AFi) exchange biased by a natural antiferromagnet like PtMn or IrMn. For the soft electrode, mostly NiFe has been used \cite{Koch98} and only recently this has been substituted by amorphous alloys of 3d ferromagnets with metalloids like B or Si, showing both, low switching fields \cite{Kaeufler02} and also high tunnel magnetoresistance (TMR) \cite{Kano02,Wang04}.

Recently, soft electrodes of polycrystalline AFis based on materials like CoFe and NiFe have been investigated. They show a reduction in stray field due to the reduced net moment, smaller switching field distribution \cite{Sousa02} and an easier establishment of a single domain structure in patterned elements with small aspect ratio \cite{Tezuka03a}.

This paper reports on the combination of both approaches, namely a soft electrode made out of two antiferromagnetically coupled layers of an amorphous CoFeB-alloy. A detailed study of the magnetic properties of this trilayer system with respect to the annealing temperature will be given. Recent results on this system have been presented, showing an oscillating antiferromagnetic (AF) coupling of $J_{\mathbf{af}} \approx 0.05$ to $0.08$ mJ/m$^2$ with the second AF maximum at a spacer thickness of 1.1nm,  and a high TMR value of approximately $50\%$ for a junction with a CoFeB-AFi as a soft magnetic electrode \cite{Wiese04a}.

Compared with a single ferromagnetic layer the AFi can be regarded as a rigid magnetic body with a reduced magnetic moment and enhanced anisotropy. The gain in coercivity can be expressed by
\begin{equation}
H_{\mathbf c}^{\mathbf {AFi}} = Q \cdot H_{\mathbf c}^{\mathbf {SL}} \label{coercivity}
\end{equation}
with $Q = (M_1 t_1 + M_2 t_2)/|M_1 t_1 - M_2 t_2|$, where $M_{\mathbf{i}}$ and $t_{\mathbf{i}}$ are the saturation magnetization and the thickness of the ferromagnetic layer {\it i}, respectively, and $H_{\mathbf c}^{\mathbf {SL}}$ and $H_{\mathbf c}^{\mathbf {AFi}}$ are the single layer and AFi coercivities \cite{Berg96}.


\section{Experimental} \label{prep}
For the ferromagnetic material of the investigated AFi we used Co$_{60}$Fe$_{20}$B$_{20}$ because of its high spin polarization, leading to high TMR values \cite{Kano02,Wang04,Wiese04a}. All samples investigated have been deposited by RF and DC sputtering on thermally oxidized SiO$_2$ wafers at a base pressure of $5 \cdot 10^{-8}$ mbar. A magnetic field of approximately 4 kA/m was applied during deposition in order to predefine the easy axis in the magnetic layers. The AFi was grown on a 1.2 nm thick Al layer, oxidized in an Ar/O${_2}$ plasma for $0.8min$ without breaking the vacuum, to have similar growth conditions as the soft electrode in a MTJ.

In order to investigate the magnetic properties of the CoFeB-based AFis and to compare them to commonly used AFis of polycrystalline material, two series of samples have been prepared. Series A consists of CoFeB(t$_1$) / Ru(1.1nm) / CoFeB(3nm) with t$_1$=3.8, 4 and 5nm, which gives nominal $Q$-values of $8.5$, $7$ and $4$, respectively. The samples with the thicker CoFeB in contact with the AlOx barrier will be called {\it positive} AFi. Additionally one {\it negative} AFi has been sputtered with CoFeB(3) / Ru(1.1) / CoFeB(3.8). Series B consists of CoFe(t$_1$) / Ru(0.9) / CoFe(t$_2$) with various thicknesses t$_1$ and t$_2$, leading to nominal Q-values in the range of 2 to 5.7. All samples were capped with a Ta layer to protect them from oxidation.

To study the temperature stability of the magnetic properties, the samples have been annealed on a hot plate at constant temperatures between $200$ and $350^{\circ}$C for $15 min.$ and have been protected from oxidation by a constant Ar-flow. The cooling times varied between 1 and 2 hours, depending on the applied annealing temperature. A field of approximately $400\frac{kA}{m}$ was applied along the easy axis during the thermal treatment.

\begin{figure}
\begin{center}
    \includegraphics[width=7cm]{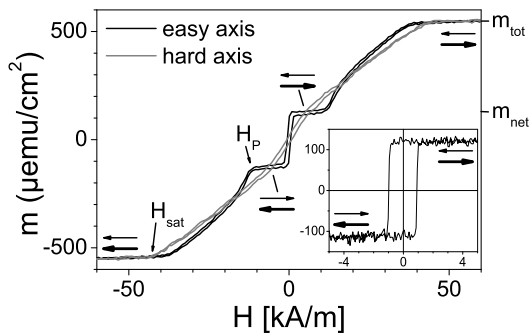}
    \caption{Magnetization loops of the AFi system Ta(5) / Al(1.2, oxid.) / CoFeB(4) / Ru(1.1) / CoFeB(3) / Ta(10) annealed at 250$^{\circ}$C.}
    \label{fig:agmloop}
\end{center}
\end{figure}

All samples then were investigated by Alternating Gradient Magnetometry (AGM). From the major and minor loops we extracted the data for the saturation field of the AFi (H$_\mathbf{sat}$), the plateau field (H$_\mathbf{P}$), the total and the net surface magnetization (m$_\mathbf{tot}$ and m$_\mathbf{net}$) and the coercivity (H$^\mathbf{AFi}_\mathbf{C}$).

\section{Results and discussion}
\begin{figure}
\begin{center}
    \includegraphics[width=6.5cm]{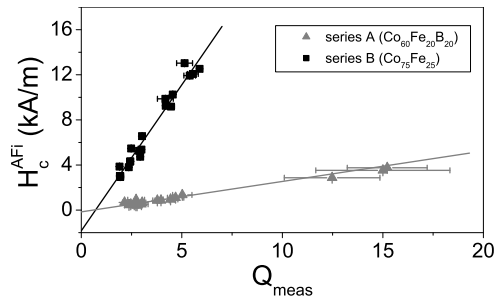}
    \caption{Coercivities of the investigated CoFeB-AFi in comparison to a CoFe-based AFi in dependance of the measured Q-values, showing much smaller coercivities for the amorphous CoFeB-trilayer. The solid lines are linear fits confirming $H_{\mathbf c}^{\mathbf {AFi}.} \propto Q$.}
    \label{fig:Hc_vs_Q}
\end{center}
\end{figure}

A typical room temperature magnetization curve, $m(H)$, of an antiferromagnetically coupled system is shown in figure \ref{fig:agmloop} after annealing at $250^{\circ}$C. The $m(H)$ curve shows a good antiparallel alignment during the magnetization reversal of the net moment of the AFi. From the minor loops (see inset) one can extract the coercivities, $H_{\mathbf c}^{\mathbf {AFi}}$, of the AF-coupled systems. When plotted against Q$_{\mathbf{meas}} = m_{\mathbf{tot}}/m_{\mathbf{net}}$, the coercivity shows a linear behavior as predicted by equation \ref{coercivity}. Therefore the coercivity can easily be tailored in a wide range by varying the Q-value of the system. Furthermore the coercivity of the CoFeB-AFis is approximately nine times smaller than in artificial ferrimagnets consisting of polycrystalline Co$_{75}$Fe$_{25}$ (see figure \ref{fig:Hc_vs_Q}), qualifying this AFi-system as an interesting material combination for soft magnetic electrodes in magnetoresistive applications.

\begin{figure}
\begin{center}
    \includegraphics[width=6cm]{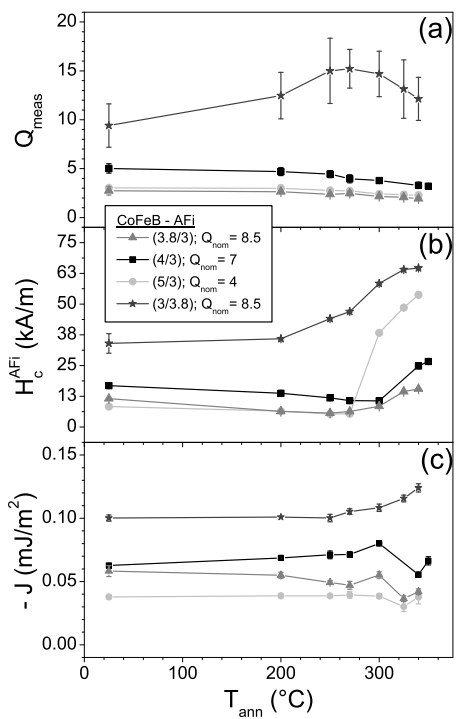}
    \caption{(a) Measured Q-value, (b) coercivity H$^{\mathbf{AFi}}_{\mathbf{c}}$ and (c) coupling $J$ of the CoFeB-trilayers against anneal temperature T$_{\mathbf{ann}}$. The values in parenthesis are the nominal layer thicknesses t$_1$, t$_2$.}
    \label{fig:temperature}
\end{center}
\end{figure}

For the samples with a {\it positive} AFi, Q$_{\mathbf{meas}}$ is significant smaller than Q$_{\mathbf{nom}}$, calculated from the nominal thicknesses of the FM layers, and is slightly decreasing with the annealing temperature [see figure \ref{fig:temperature}(a)]. This discrepancy between Q$_{\mathbf{meas}}$ and Q$_{\mathbf{nom}}$ can be explained by thicker magnetically dead layers of the upper CoFeB-layer in comparison to the bottom layer, as extracted also from the $m(H)$ loops. Since the samples are well protected from oxidation, as confirmed by Auger depth profiling, this indicates a stronger intermixing of the upper CoFeB interfaces, that leads to a increase of $m_{\mathbf{net}}$ and therefore a decrease of Q$_{\mathbf{meas}}$.

For the same reason the {\it negative} AFi of series A shows a much higher Q$_{\mathbf{meas}}$ in comparison to the nominal Q of 8.5. Additionally a strong increase of Q-value for temperatures up to 270$^\circ$C is found. This increase is also reflected by an enlargement of coercivity in this temperature range, as expected by equation \ref{coercivity}, whereas the increase at T$_{\mathbf{ann}}>300^\circ$C is also seen in the {\it positive} samples of series A. For those the coercivity slightly decreases in the temperature range of $200$ to $300^{\circ}$C and for temperature higher than $300^{\circ}$C it strongly increases [figure\ref{fig:temperature}(b)] and most likely indicates the onset of the change from amorphous to polycrystalline phase. This behavior is also seen for tunnel junctions using a single layer of CoFeB as the soft electrode at this temperature \cite{Dimopoulos_JAP04}.


The extracted values of the saturation field, $H_{\mathbf {sat}}$, the total, $m_{tot}=m_1+m_2$, and the net surface magnetization, $m_{net}=m_1-m_2$, allows one to calculate the individual magnetization of the layers, m$_{1,2}$ ,and the coupling energy ${J = - \mu_0 H_{\mathbf {sat}} \frac{m_1 m_2}{m_1 + m_2}}$ \cite{Berg97}. As can be seen in figure \ref{fig:temperature}(c), the coupling is already set in the as deposited state for all samples and remains stable for the {\it positive} stack system until 300$^\circ$C. For these samples the coupling first decreases above 300$^\circ$C, and at temperatures above $340^\circ$C finally the plateau, defined by H$_\mathbf{P}$, disappears. This is maybe due to a transition from a bilinear to a dominating biquadratic coupling, which most likely originates from thickness variations in the Ru-layer due to interdiffusion.

$J_{\mathbf{af}}$ for the {\it positive} samples varies between $0.037$ and $0.063$ mJ/m$^2$, whereas the {\it negative} AFi shows a coupling of $0.12$ mJ/m$^2$. This discrepancy can be explained by the fact, that the magnetic dead layer thickness is larger for the upper FM layer. This leads for the case of the {\it negative} AFi to a decrease of its net moment with a consequently increase of the saturation field and of $J_{\mathbf{af}}$.
Nevertheless the coercivity of the AFi systems is independent from the absolute value of coupling, as long the two magnetic layers are antiferromagnetically oriented in the operational window. This is confirmed by the linear behavior of coercivity vs. Q-value in figure \ref{fig:Hc_vs_Q}.


\section{Summary} \label{summary}
We have shown, that CoFeB/Ru/CoFeB sandwiches exhibit a stable coupling up to T$_{\mathbf{ann}} \approx 325^\circ$C. At the same time the coupling energy $J$ is in the order of 0.1 mJ/m$^2$ and therefore by a factor of ten smaller than in polycrystalline CoFe/Ru/CoFe AFis. The coercivity of the amorphous AFi is by a factor of nine smaller than in the polycrystalline samples and scales linearly with the measured Q-value. Combined with the high spin polarization of the CoFeB alloy, reflected by a measured TMR effect of approximately 50$\%$, this material system may be a potential candidate as a soft magnetic electrode in magnetic tunnel junctions.

The authors wish to thank J. Bangert and G. Gieres for fruitful discussions, and H. Mai for experimental support. Financial support of the German Ministry for Education and Research is gratefully acknowledged (grant 13N8208).







\end{document}